\documentclass[10pt,twoside]{classe-cmb}
\usepackage{amssymb,amsbsy,amsmath,amsfonts,amssymb,amscd}
\usepackage{latexsym,euscript,exscale,epic,eepic,epsfig}
\usepackage[english,francais]{babel}
\usepackage{times}
\usepackage{referj}

\newtheorem{e-proposition}[theorem]{Proposition}

\newtheorem{e-definition}[theorem]{Definition\rm}

\newcommand{\ETAL}{et al.\ }

\newcommand{\Npix}{N_{\rm pix}}

\newcommand{\fbP}{\delta T_{\rm fb}}

\newcommand{\vfbP}{\overrightarrow{\delta T_{\rm fb}}}

\newcommand{\sigN}{\sigma_N}

\newcommand{\vd}{\overrightarrow{d}}

\newcommand{\vTheta}{\overrightarrow{\Theta}}

\newcommand{\vTeta}{\overrightarrow{\Theta}}

\newcommand{\mC}{\boldsymbol{C}}
\newcommand{\mT}{\boldsymbol{T}}

\newcommand{\mN}{\boldsymbol{N}}

\newcommand{\mW}{\boldsymbol{W}}
\newcommand{\mM}{\boldsymbol{M}}

\newcommand{\vTetab}{\overrightarrow{\Theta}_{best}}

\newcommand{\C}{{\boldsymbol {\cal C}}}
\newcommand{\calN}{{\cal N}}

\ComParit{S1296-2147}
\PIT{FLA}
\PXHY{????}
\Add{?}
\Volume{0}
\Year{2003}
\FirstPage{1}
\LastPage{??}
\AuteurCourant{..}
\TitreCourant{..} %less than 60 characters
\Journal
\Rubrique{Rubrique}{Heading}
\SousRubrique{Sous-rubrique}{Sub-Heading}
\PresentePar{First name}{NAME}
\Recu{blank}{}{}
\begin{document}
\TitleOfDossier{The Cosmic Microwave Background: \\present status and
 cosmological perspectives}
\title{Cosmological Parameter Estimation: Method%The article title
}
\author{%
Marian Douspis~$^{\text{a}}$ %,\ \
%Second author~$^{\text{b}}$,\ \
%Third author~$^{\text{c}}$
}
\address{%
\begin{itemize}\labelsep=2mm\leftskip=-5mm
\item[$^{\text{a}}$]
Astrophysics, Oxford University, Keble Road, OX1 3RH, Oxford, UK\\
E-mail: douspis@astro.ox.ac.uk
%\item[$^{\text{b}}$]
%Second address \\
%E-mail:
%\item[$^{\text{c}}$]
%Third address \\
%E-mail:
\end{itemize}
}
\maketitle
\thispagestyle{empty}
%%%%%%%%%%%%%%%%%%%%%%%%%%%%%%%%%%%%%%%%%%%%%%%%%%%%%%%%%%%%
%%%  Abstract  %%%
%%%%%%%%%%%%%%%%%%
\begin{Abstract}{%
 CMB anisotropy data could put powerful constraints on theories of the
 evolution of our Universe.  Using the observations of the large
 number of CMB experiments, many studies have put constraints on
 cosmological parameters assuming different frameworks. Assuming for
 example inflationary paradigm, one can compute the confidence
 intervals on the different components of the energy densities, or
 the age of the Universe, inferred by the current set of CMB
 observations. The aim of this note is to present some of the
 available methods to derive the cosmological parameters with their
 confidence intervals from the CMB data, as well as some practical
 issues to investigate large number of parameters.  }\end{Abstract}
%\par\medskip\centerline{\rule{2cm}{0.2mm}}\medskip
\setcounter{section}{0}
\begin{Abstract}{%
 Les observations des anisotropies du fond diffus cosmologique (FDC)
 peuvent placer de fortes contraintes sur les th\'eories d'\'evolution
 de notre Univers. L'utilisation de telles donn\'ees a permis de
 contraindre differents param\`etres de diff\'erents cadres
 th\'eoriques: l'age de l'Univers, son contenu baryonique, etc. Le but
 de cette contribution est de pr\'esenter differentes m\'ethodes
 possibles pour extraire les param\`etres cosmologiques et leurs
 intervalles de confiance des donn\'ees du FDC. Des questions
 pratiques sur l'utilisation de grands nombres de param\`etres sont
 aussi abord\'ees.}\end{Abstract}
\par\medskip\centerline{\rule{2cm}{0.2mm}}\medskip
\setcounter{section}{0}
\selectlanguage{english}
%%%%%%%%%%%%%%%%%%%%%%%%%%%%%%%%%%%%%%%%%%%%%%%%%%%%%%%%%%%%
%%%  Main text (in English)  %%%
%%%%%%%%%%%%%%%%%%%%%%%%%%%%%%%%

%%%%%%%%%%%%%%%%%%%%%%%%%%%%%%%%%%%%%%%%%%%%%%%%%%%%%%%%%%%%% 1. %%%%%%%%%%%%
\section{Introduction}

        The extraction of information from cosmic microwave
background (CMB) anisotropies is a classic problem of
model testing and parameter estimation, the goals
being to constrain the parameters of
an assumed model and to decide if
the {\em best--fit model} (parameter values)
is indeed a good description of the data.
Maximum likelihood is often used as the method
of parameter estimations.  Within the context
of the class of models to be examined, the probability
distribution of the data
%for an arbitrary set of parameters
is maximized as a function of the model parameters, given the actual,
observed data set\footnote{In recent Bayesian analyses, quoting the
mean of the product of the likelihood and prior functions as best
model, is preferred}.
%This is the same as a Bayesian analysis with uniform priors.
 Once found, the best model must then be judged on its
ability to account for the data, which requires the construction of a
{\em statistic} quantifying the {\em goodness--of--fit} (GoF).
Finally, if the model is retained as a good fit, one defines {\em
confidence} intervals on the parameter estimation.  The exact meaning
of these confidence intervals depends heavily on the method used to
construct them, but the desire is always the same -- one wishes to
quantify the `ability' of other parameters to explain the data (or
not) as well as the best fit values. Given the quality of the current
data, and the aim of the analysis -- precise determination of the
cosmological parameters -- much attention should be put on the
robustness and accuracy (unbiased techniques) of the methods used.  I
review the different ways of estimating the likelihood function of the
parameters focusing on the use of the angular power spectrum
($C_\ell$'s). Then, some methods to compute the goodness of fit and
the confidence intervals will be discussed. Finally, some practical
issues for such computations will be addressed. In this review, I take
the temperature fluctuations as the observed quantity. The same
approaches could be applied for the polarisation signal of the CMB.

%%%%%%%%%%%%%%%%%%%%%%%%%%%%%%%%%%%%%%%%%%%%%%%%%%%%%%%%%%%%% 2. %%%%%%%%%%%%
\section{Likelihood}

  Data on the CMB consists of sky brightness measurements, usually
  given in terms of equivalent temperature in pixels.  The likelihood
  function is to be constructed using these pixel values\footnote{The
  term pixel will be understood to also include temperature
  differences.}.  Standard Inflationary scenarios predict {\em
  Gaussian} sky fluctuations, which implies that the pixels should be
  modeled as random variables following a multivariate normal
  distribution, with covariance matrix given as a function of the
  model parameters (in addition to a noise term). It is important to
  note that, since the parameters enter through the covariance matrix
  in a non--linear way,the likelihood function ${\cal L}$ is not a
  linear function of the (cosmological) parameters.

  Although it would seem straightforward to estimate model parameters
  directly with the likelihood function from the maps ({\em full
  analysis}), in practice the procedure is considerably complicated by
  the complexity of the model calculations and by the size of the data
  sets (\cite{bond00a, bond00b, borril99b, kogut}).  Maps consisting
  of several hundreds of thousands of pixels (the present situation)
  are extremely cumbersome to manipulate, and the million--pixel maps
  expected from Planck cannot be analyzed by this method in any
  practical way.  An alternative is to first estimate the angular
  power spectrum from the pixel data and then work with this reduced
  set of numbers.  For Gaussian fluctuations, there is in principle no
  loss of information.  Because of the large reduction of the data
  ensemble to be manipulated, the tactic has been referred to as
  ``radical compression'' (\cite{bond00b}).  The power spectrum has in
  fact become the standard way of reporting CMB results; it is the
  best visual way to understand the data, and in any case it is what
  is actually calculated in the models.  The first part of this
  section describes briefly the full analysis procedure. Then, the
  second part will focus on the power spectrum as starting point for
  cosmological parameters estimation. In the latter case, due to the
  non-Gaussian behavior of the $C_\ell$'s, elaborated approximations
  should be used.

%%%%%%%%%%%%%%%%%%%%%%%%%%%%%%%%%%%%%%%%%%%%%%%%%%%%%%%%%%%%% 2.1 %%%%%%%%%%%%
\subsection{Full analysis}

 Temperature fluctuations of the CMB are described by a random field
 in two dimensions: $\Delta(\hat{n})\equiv (\delta T/T) (\hat{n})$,
 where $T$ refers to the temperature of the background and $\hat{n}$
 is a unit vector on the sphere.  It is usual to expand this field
 using spherical harmonics: 
 \begin{equation} \label{eq:alm}
 \Delta(\hat{n}) = \sum_{\ell m} a_{\ell m} Y_{\ell m}(\hat{n})
 \;\;\;\; <a_{\ell m}a^*_{\ell'm'}>_{ens} = C_\ell
 \delta_{\ell\ell'}\delta_{mm'} 
 \end{equation} 
 The $a_{\ell m}$'s are randomly selected from the probability
 distribution characterizing the process generating the
 perturbations. In the Inflation framework, which we consider here,
 the $a_{\ell m}$'s are {\em Gaussian random variables} with zero mean
 and covariance\footnote{The indicated averages are to be taken over
 the theoretical ensemble of all possible anisotropy fields, of which
 our observed CMB sky is but one realization.} given in Eq.~1.  The
 $C_\ell$'s then represent the {\em angular power spectrum}.  We may express
 the observed (or beam smeared) correlation between two points
 separated on the sky by an angle $\theta$ as
 \begin{equation} \label{eq:ctheta}
 C_b(\theta) \equiv <\Delta_b(\hat{n}_1)\Delta_b(\hat{n}_2)>_{ens} =
 \frac{1}{4\pi}\sum_{\ell} (2\ell+1) C_\ell B_\ell^2 P_\ell(\mu)
 \end{equation} 
 where $P_\ell$ is the Legendre polynomial of order $l$, $\mu =
 \cos\theta = \hat{n}_1\cdot\hat{n}_2$ and $B_\ell$ is the
 harmonic coefficient of the beam decomposition\footnote{Note that
 this expansion pre-supposes axial symmetry for the beam}. The statistical
 isotropy of the perturbations demands that the correlation function
 depend only on separation, $\theta$, which is in fact what permits
 such an expansion.

 Given these relations and a CMB map, it is now straightforward to
 construct the likelihood function, whose role is to relate the
 $\Npix$ observed sky temperatures, which we arrange in a {\em data
 vector} with elements $d_i \equiv \Delta_b(\hat{n}_i)$, to the model
 parameters, represented by a {\em parameter vector} $\vTeta$.  For
 {\em Gaussian} fluctuations (with Gaussian noise) this is simply a
 multivariate Gaussian:
 \begin{equation}\label{eq:like1} {\cal L}(\vTeta) \equiv {\rm
 Prob}(\vd|\vTeta) = \frac{1}{(2\pi)^{\Npix/2} |\mC|^{1/2}}
 e^{-\frac{1}{2} \vd^t \cdot \mC^{-1} \cdot \vd} 
 \end{equation}
 The first equality reminds us that the likelihood function is the
 probability of obtaining the data vector given the model as defined
 by its set of parameters.  In this expression, $\mC$ is the pixel
 covariance matrix: 
 \begin{equation}\label{eq:covmat} C_{ij} \equiv
 <d_id_j>_{ens} = T_{ij} + N_{ij} 
 \end{equation} 
 where the expectation value is understood to be over the theoretical
 ensemble of all possible universes realizable with the same parameter
 vector.  The second equality separates the model's pixel covariance,
 $\mT$, from the noise induced covariance, $\mN$.  According to
 Eq.~\ref{eq:ctheta}, $T_{ij} = C_b(\theta_{ij})\equiv 1/(4\pi)
 \sum_\ell (2\ell+1) C_\ell W_{ij}(\ell)$ where $\mW$, the window  
 matrix, contains the beam and strategy effects (direct measure,
 differences).  The parameters may be either the individual $C_\ell$
 (or band--powers, discussed below), or the fundamental cosmological
 constants, $\Omega, H_o$, etc...  
 In the latter situation, the parameter dependence enters through
 detailed relations of the kind $C_\ell[\vTeta]$, specified by the
 adopted model (e.g., Inflation).

  For cosmological parameters estimations, one has to compute the
  likelihood value of Eq.~\ref{eq:like1} for a family of models
  investigated. For each set of parameters $\vTheta$, the
  computational time for the likelihood goes like $\Npix^3$, unless
  geometrical symmetries in the observational strategy allows to use
  faster algorithm for inverting $\mC$. Investigating one handful of
  parameters with reasonable steps and ranges (typically
  $N_{parameters}^{10}$) with a map of few thousands of pixels becomes
  extremely cumbersome. Only few studies has been done in such a way
  (\cite{ratra, rocha,gorski,DBBL}. Such a computation with
  second generation experiments is thus prohibitive\footnote{except
  for some particular symmetries, \cite{wandelt2}}.

%%%%%%%%%%%%%%%%%%%%%%%%%%%%%%%%%%%%%%%%%%%%%%%%%%%%%%%%%%%%% 2.2 %%%%%%%%%%%%
\subsection{Using the Angular Power Spectrum}

\begin{figure*}[!t]
\begin{center}
\resizebox{!}{!}{\includegraphics[totalheight=8cm,
        width=10cm]{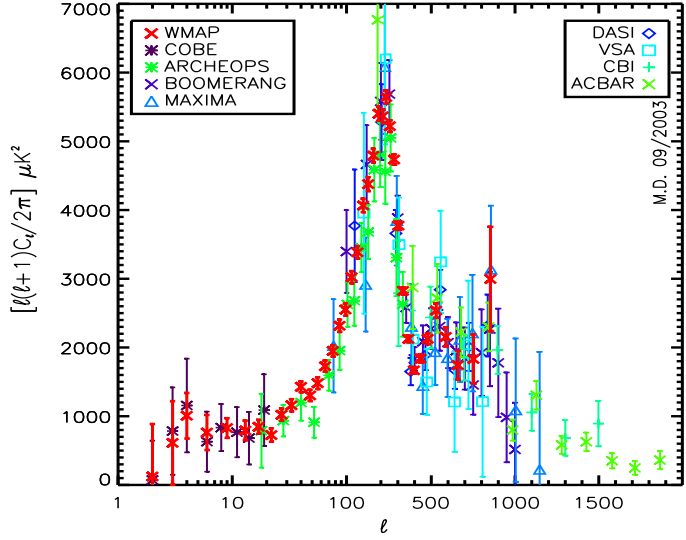}}
\end{center}
\caption{\label{PS}Angular power spectrum estimates of the CMB anisotropies in September 2003 (\cite{wmap, cobe, archeops, boom2, maxima, dasi,vsa, cbi, acbar}). Notice the good agreement between  band powers coming from different experiments (different detectors, technology, scanning strategy,...) until $\ell\sim 500$.}
\end{figure*}

 In order to avoid the problem of computational cost of the full
 analysis, an alternative consists in first estimating the angular
 power spectrum from the pixel data and then work with the latter to
 estimate the cosmological parameters.  The critical issue is then how
 to correctly use the power spectrum for an unbiased parameter
 estimation and model evaluation.  The angular power spectrum can be
 evaluated with different techniques (see Hamilton, this issue,
 \cite{hamilton}). Again, a likelihood analysis from the maps can be
 done by inserting a spectral form into the definition of $\mT$.  For
 example, the commonly used {\em flat} band--power, $\delta T_{fb}$
 (or $\C_b = \delta T_{fb}^2$ over a certain range in $\ell$),
 actually represents the equivalent logarithmic power integrated over
 the band, which simplify the correlation matrix as follows:
\begin{equation}\label{eq:dtfb}
 C_\ell \equiv 2 \pi [\delta T_{fb}^2 / (\ell (\ell +1)] \;\;\;\;\;\;
 \mT = \frac{1}{2} \fbP^2 \sum_\ell \frac{(2\ell+1)}{\ell(\ell+1)}
 \mW(\ell) \end{equation} In this way, we may write Eq.~\ref{eq:like1}
 in terms of the band--power and treat the latter as a parameter to be
 estimated.  This then becomes the band--power likelihood function,
 ${\cal L}(\fbP)$. Figure~\ref{PS} shows the latest band power
 estimates of the CMB fluctuations. Some of the points have been
 obtained by maximizing this likelihood function; the errors are
 typically found by in a Bayesian approach, by integration in $\C_b$
 over ${\cal L}$ with a uniform prior (eg. DASI\cite{dasi},
 VSA\cite{vsa}, CBI\cite{cbi}, ACBAR\cite{acbar}). Other band powers
 and errors are estimated by using Monte Carlo based methods (see
 \cite{master, szapudi}) like the WMAP\cite{wmap},
 BOOMERANG\cite{boom2}, MAXIMA\cite{maxima} and
 Archeops\cite{archeops} ones. Notice that the variance due to the
 finite sample size (i.e., the sample variance, including the cosmic
 variance due to our observation of one realization of the sky) is
 fully incorporated into the analyses.

% -- the likelihood function ``knows'' how many pixels there are.  
% Once again, the flat band power estimates represent the 
%{\em variance} of Gaussian distributed pixel values (the sky 
%temperature fluctuations), and they therefore do not have a 
%Gaussian probability distribution, and thus the likelihood isnot a 
%Gaussian function.

 Given a set of band--powers
 how should one proceed to constrain the fundamental cosmological
 parameters, denoted in this subsection by $\vTeta$?  If we had an
 expression for ${\cal L}(\vfbP)$, for our set of band--powers
 $\vfbP$, then we could write ${\cal L}(\vfbP) = Prob(\vd|\vfbP) =
 Prob(\vd|\vfbP[\vTeta]) = {\cal L}(\vTeta)$.  Thus, our problem is reduced to
 finding an expression for ${\cal L}(\vfbP)$, but as we have seen,
 this is a complicated function of $\vfbP$, requiring use of all the
 measured pixel values and the full covariance matrix with noise --
 the very thing we are trying to avoid.  Our task then is to find an
 approximation for ${\cal L}(\vfbP)$.

%%%%%%%%%%%%%%%%%%%%%%%%%%%%%%%%%%%%%%%%%%%%%%%%%%%%%%%%%%%%% 2.2.1 %%%%%%%%%%
\subsubsection{$\chi^2$ minimization}

The most obvious way of finding ``the best model'' given a set of
points and errors is the traditional $\chi^2$--minimization. This means 
that we assume a Gaussian shape for the likelihood function of the kind:
\begin{equation}\label{eq:chi2c}
{\cal L}(\vfbP) = e^{-\chi^2(\vTheta)/2.}= e^{-(\vd^{obs}-\vd^{model}) \cdot \mM^{-1} \cdot (\vd^{obs}-\vd^{model})/2.}
\end{equation}
where $\mM$ is the correlation matrix between the different flat band
estimates.  The main problem with this approach is that it deals with
the flat band--power estimates as Gaussian distributed data which they
are not (obeying the statistics of the square of a Gaussian).  Then,
it has been shown that such a procedure gives a biased estimation of
the cosmological parameters and bad estimates of the confidence
intervals (\cite{bond00b,BDBL}), leading to the search of more
accurate approximation of the likelihood function.

%%%%%%%%%%%%%%%%%%%%%%%%%%%%%%%%%%%%%%%%%%%%%%%%%%%%%%%%%%%%% 2.2.2 %%%%%%%%%%
\subsubsection{More elaborated approximations}

Different studies have been made to reconstruct better analytical
approximations directly from the form of the flat band likelihood
function (\cite{bond00b,BDBL, wandelt}). This section will focus on
two of them. One is derived from the likelihood function in a
particular case, for which it is actually exact \cite{BDBL}. The other
one, mostly used during the last years, offers the advantage of being
really close to a $\chi^2$ minimization by changing variables in the
appropriate way \cite{bond00b}.  Both approximations need a small
amount of information and aim to be used directly from the spectrum
given in the literature\footnote{As we will see these approximations
need one more information than the basic $\chi^2$ minimization in
order to take into account the non Gaussian behavior of the likelihood. The
authors have been asking that this information is provided in addition
to the band powers estimates and errors. Recent experiments have
published the necessary information in their papers.}.

\begin{itemize}
\item BDBL approximation
\end{itemize}

\begin{figure*}[!t]
\begin{center}
\resizebox{\hsize}{!}{\includegraphics[totalheight=10cm,
        width=16cm]{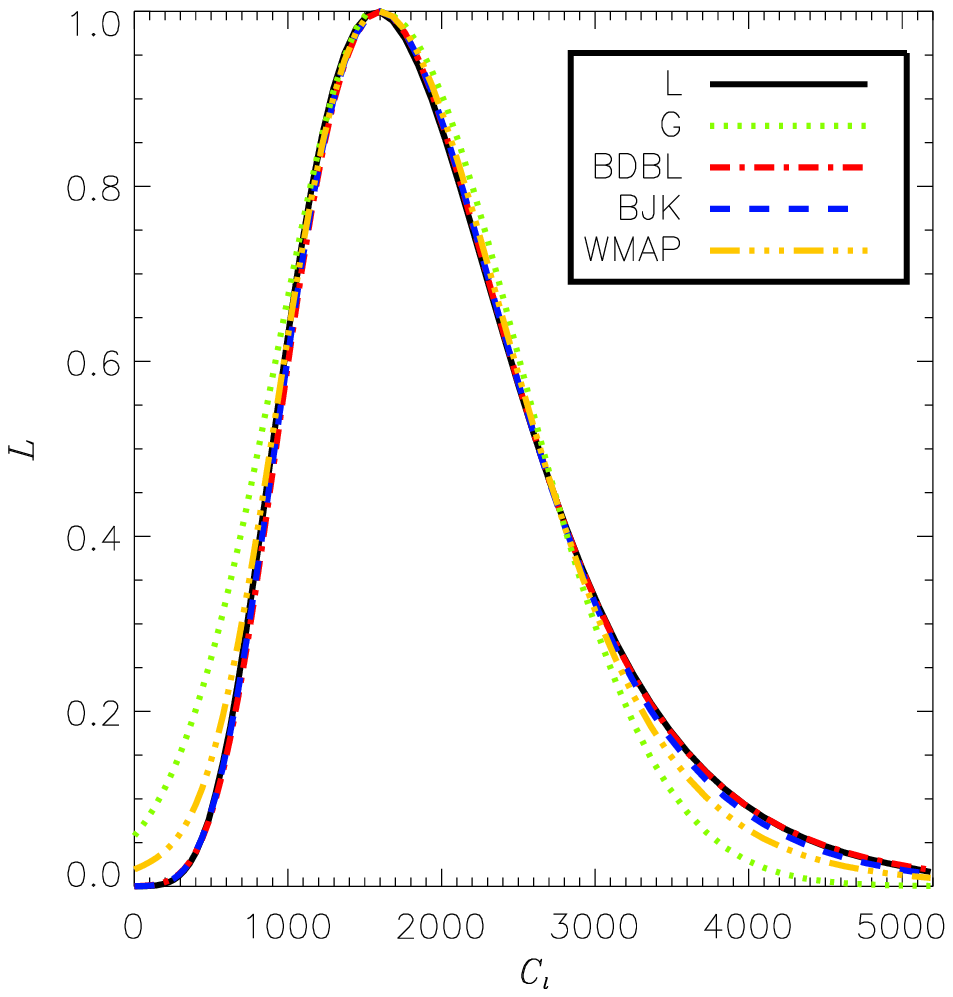}\includegraphics[totalheight=10cm,
        width=16cm]{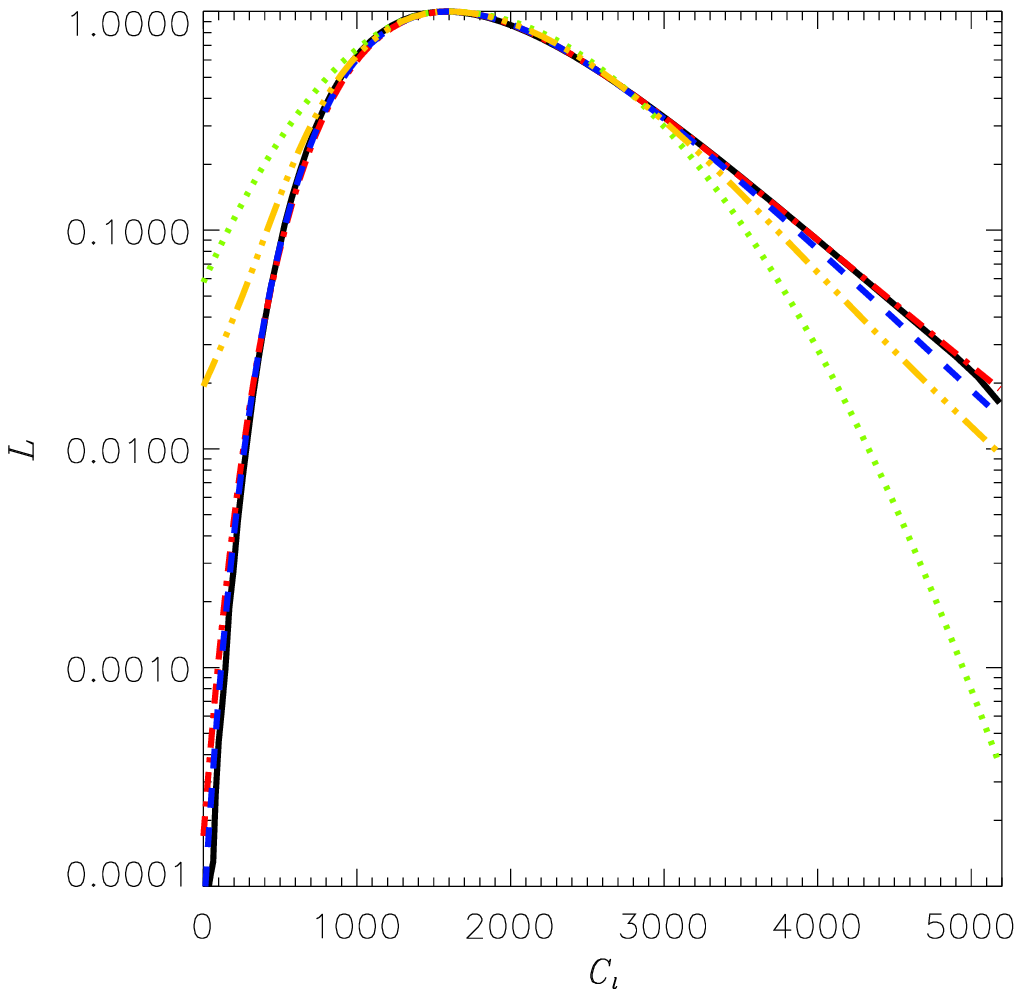}}
\end{center}
\caption{\label{fig_approx}Comparison to the TOCO97 likelihood function for all approximation described in this section. The black solid line shows the full likelihood function computed from the map. The red dot--dashed and blue dashed lines show the BDBL and BJK approximation respectively. The 2--wings Gaussian and WMAP--type approximation are plotted in dotted green and dot-dot-dashed yellow lines}
\end{figure*}

The Bartlett, Douspis, Blanchard and Le Dour approximation is based on
the analytical form of the likelihood in an ``ideal'' experiment,
where all the pixels ($\Npix$) are independent random variables
(uncorrelated) and the noise is uncorrelated and uniform ($\sigN^2$).
In that particular case, one can write the exact likelihood function
as follows:
\begin{equation}\label{eq:approx}
{\cal L}(\fbP)  \propto  X^{\nu/2} e^{-X/2} \;\;\;\;\; X[\fbP] \equiv  \frac{([\fbP^{(o)}]^2 + \beta^2)}{([\fbP]^2 +
        \beta^2)}\nu 
\end{equation}
where $\beta=\sigN$ and $\nu=\Npix$. The approximation comes from the
fact that the authors keep the same likelihood form for real
experiments, whereas the noise is no longer uniform and uncorrelated,
and the pixels are not independent. To take into account these
differences, one lets $\nu$ and $\beta$ as free parameters and fixes
them by fitting the 68\% and 95\% confidence intervals (published or
inferred from the true likelihood function). 
Figure \ref{fig_approx} shows the comparison between this
approximation and true likelihood functions obtained for TOCO data
(\cite{toco, tocoweb}).

The advantage of this approximation is that the better the behavior
of the experiment is (less correlations, more uniform noise, ...), the
better the approximation is; being exact for the ideal case. It is
also unbiased at the maximum of the likelihood and allows one to
recover the full shape of the likelihood function. The inconvenient of
this approximation is that it is defined only for uncorrelated flat
band powers. The possible correlations between bands are not taken
into account, as the full likelihood is given by the product of all
individual likelihood functions: ${\cal L}(\vfbP) = \prod {\cal
L}(\fbP^i)$.

\begin{itemize}
\item BJK approximation
\end{itemize}

In the second case, also referred as the Bond, Jaffe \& Knox
approximation the motivation is driven by the need to work with
Gaussian distributed variables for which the $\chi^2$ is not biased
any more. Writing the likelihood in the spherical harmonic space for
the same ideal experiment as above, and considering
$\C_\ell=\ell(\ell+1)C_\ell/(2\pi)=\delta T^2_\ell$,
$\calN_\ell=\ell(\ell+1)N_\ell/(2\pi)$ where $N_\ell=\langle|n_{\ell
m}|^2\rangle$ is the noise power spectrum in spherical harmonics, one
can show that the curvature matrix evaluated at the maximum is
proportional to
$\left(\C_\ell+\calN_\ell/B_\ell^2\right)^{-2}\delta_{\ell\ell'}$.  If
one define $Z_\ell \equiv \ln(\C_\ell + x_\ell)$ where in this
particular ``ideal experiment'' $x_\ell=\calN_\ell/B_\ell^2$, the
curvature matrix expressed in term of $Z$ is then constant. BJK
approximation to the likelihood is then to take $Z_b$ (determined in a
band) as normally distributed in realistic experiments (by finding the
good expression of the corresponding $x_b$).  From the previous statements,
one can then express the likelihood by:
\begin{equation}\label{eqn:gausslike2}
 {\cal L}(\vfbP)= e^{(-Z\cdot \mM^{-1}\cdot Z^t)/2.} \;\; {\rm where}\;\; Z_i = \ln (\fbP^2(i) + x_b(i))\;\; {\rm evaluated\; in\; a\; band}\;i
\end{equation}    
 The absolute value of Eq.~\ref{eqn:gausslike2} gives also an estimate
 of the goodness of fit. As we will see below, this approximation is
 slightly biased at the maximum of the likelihood but has been shown
 to be a reliable approximation (see Fig.~\ref{fig_approx}) and is
 available online through the RADPACK package of .;\cite{radpack}.\\

The WMAP team \cite{verde} adopted an hybrid approximation: $\ln {\cal
L} = \frac{1}{3} \ln {\cal L}_{Gauss} + \frac{2}{3} \ln {\cal
L}_{BJK}$ motivated by an expansion of the true likelihood around the
maximum\footnote{where ${\cal L}_{Gauss}= \exp(-\chi^2/2)$}. This
formulation has the advantage to be unbiased around the maximum but
has not been tested against the real likelihood function in the wings.

%Figure~\ref{fig_approx} shows the difference of shape of the different
%approximation considered in this section.

Once the likelihood function ${\cal L}(\vfbP)$ is known, one is able
to compute ${\cal L}(\Theta)$ for a family of models. As we have seen,
temperature on the sky are random Gaussian variables and then the
``radical compression'' is thus valid and induces no loss of
information. The latter is true only if all the spectrum (in the limit
of sensitivity of the experiment; the window function) is
specified. Whereas, for different reasons (partial sky coverage, noise
correlation, ...) only the spectrum {\em in band} is recovered: the
spectrum is approximated by steps in $\ell$. Such description induces
a loss of information which may have some effect on the cosmological
parameter estimation (bias and degeneracies). Douspis et al. (\cite{DBBL})
have shown that a better description of the spectrum (power in band
{\em and} slope in band) could decrease the bias. The second and third
generation of experiments provides (or will provide) better
sensitivity, less correlated measurements which allows one to recover
the spectrum with better resolution in $\ell$ (see WMAP for ex.),
decreasing therefore the bias.  Most of the studies are nevertheless
performed by using the set of flat band estimates likelihood
functions.

%%%%%%%%%%%%%%%%%%%%%%%%%%%%%%%%%%%%%%%%%%%%%%%%%%%%%%%%%%%%% 3. GOF %%%%%%%%%%
\section{Goodness of fit}

Once the (approximated) likelihood values of the models investigated
are computed, one should find the best model (maximum) and  evaluate the
quality of the fit before constructing the parameter constraints. As a
general rule, one must judge the quality of the fit before any serious
consideration of the confidence intervals on parameters. This
requires the application of a {\em Goodness of fit} (GOF)
statistic. The latter is usually a function of both the model and the
data, which reaches a maximum (or minimum) when the data is generated
from the theory. The 'significance' may then be 
defined as the probability of obtaining $gof > gof_{\rm obs}$.
On this basis, it permits a quantitative evaluation of 
the quality of the best model's fit to the data: if the probability
of obtaining the observed value of the GOF statistic (from
the actual data set) is low (low significance), then
the model should be rejected.  Without such a statistic, one does 
not know if the best model is a good model, or simply 
the ``least bad'' of the family.

In the full likelihood analysis method, the best model (set of
parameters) could be obtained by maximizing the likelihood function of
Eq~\ref{eq:like1} and is defined by $\vTetab$ in the following. One
can easily note the Gaussian form of Eq~\ref{eq:like1} in the data
vector $\vd$. Given the best model, the most obvious GOF statistic is
then clearly $gof = \vd^t \cdot \tilde{\mC}^{-1} \cdot \vd $ where
$\tilde{\mC} \equiv C(\vTetab)$ is the correlation matrix evaluated at
the best model. For the Gaussian fluctuations we have assumed, this
quantity follows a $\chi^2$ distribution, with a number of
degrees--of--freedom (DOF) approximately equal to the number of pixels
minus the number of parameters\footnote{This recipe does not strictly
apply in the present case, because the parameters are non--linear
functions of the data; it is nevertheless standard practice.  In any
case, the number of pixels is in practice much larger than the number
of parameters. The numbers of degrees of freedom is also less than the
number of pixels because of the correlations between pixels (non
diagonal correlation matrix). Nevertheless, the matrices are mainly
diagonal and the $gof$ is then mostly insensitive to the small
reduction of the number of DOF.}.

The use of $\chi^2$ method (in $\fbP$ or any change of variables like
in BJK approximation) makes even easier the computation of the
GOF. The obvious GOF statistics would just be one number, the value of
the $\chi^2$ evaluated at the minimum: $gof = \chi^2(\vTetab)$. It is
of course true that if the number of contributing effective
DOF is large, a power estimate will closely follow a
Gaussian; this, however, is never the case on the largest scales
probed by a survey.  Douspis et al., \cite{GOF}, have shown for
example that the $\chi^2$ approach leads to quantitatively different
results than other, more appropriate GOF statistics.

 When more elaborated approximations are used, the goodness of fit
 computation is less obvious.  One should first reconstruct the
 distribution of the estimators. This could be a natural output when
 Monte Carlo based methods are used for the $C_\ell$'s extraction
 \cite{master, szapudi}, but it is mostly unknown when one applies
 traditional methods. Douspis et al. \cite{GOF} have proposed an approximation
 which allows to reconstruct the distribution from the shape of the
 flat--band likelihood function\footnote{This technique could be used
 both ways, allowing to reconstruct the likelihood when the
 distribution is known}. When the latter is known, one should build a
 GOF statistique in order to compute to data probability given the
 best model (see \cite{GOF} for examples).

Knowing that the best model is indeed a good fit to the data, or that
the data have a good chance to be generated from this model, one
should proceed by estimating the confidence intervals on the
investigated parameters.

%%%%%%%%%%%%%%%%%%%%%%%%%%%%%%%%%%%%%%%%%%%%%%%%%%%%%%%%%%%%% 4.  CI %%%%%%

\section{Confidence Intervals}

The estimation of confidence intervals is mostly a question of
definition. Most of CMB analyses have been done in the Bayesian
framework and are thus dependent on the priors assumed. Some
frequentist attempts have been performed in order to eliminate such
dependencies. The reader can read more about the comparison between the
two methods in \cite{abroe}. 

Typically, the frequentist analyses are related to the
goodness-of--fit statistics and the probability distribution of data
for a given model (set of parameters). 
In the Bayesian approach, one reconstructs the conditional (posterior)
probability density function (pdf), $P(\vTeta_{true}|d_{obs})$, for
the unknown $\vTeta_{true}$ given the observation
$d_{obs}$\footnote{In our problem, $d_{obs}$ should be taken as a set
of flat band powers}, from the pdf (which dependency in $\vTeta$ is
known) for observing $d$ using Bayes theorem. The latter evaluated at
$d_{obs}$ is known as Likelihood function: $P(d_{obs}|\vTeta)={\cal
L}(d_{obs}|\vTeta)$.
\begin{equation} 
P(\vTeta_{true}|d_{obs})={\cal L}(d_{obs}|\vTeta_{true}) P(\vTeta_{true})/P(d_{obs})
\end{equation} 
The denominator is just a normalization factor, and thus one of the
issues is what to use for the prior $P(\vTeta_{true})$. If one knows
the likelihood and fixes the prior (usually taken as uniform in terms
of the parameters) then one knows the posterior probability
distribution. A Bayesian credible region (interval) for a parameter is
the range of parameter values that encloses a fixed amount of such
probability. As the questions asked in the two approaches are quite
different, one does not expect necessarily the intervals computed in
the two methods to be similar.

I will described in the following two ways of estimating the
confidence intervals referred as marginalisation and maximization.
These two approaches are usually presented in opposition.
% but the better behavior of the likelihood may have, the most similar the
%results are. 
In the limit of a Gaussian shaped likelihood with linear dependencies
in the parameters, the two techniques are equivalent (see
demonstration in \cite{tegmark}). Unfortunately this is not the case
in cosmological parameter estimation. Both techniques consist in two
steps: first one reduces the number of parameters in order to
visualize the likelihood (or pdf) function (or surface in 2
dimensions). Then one computes the confidence intervals for each
parameter.

\subsection{Marginalisation}

In CMB analyses, as $\vTeta$ is usually a vector of 5 to 10
parameters, it is quite hard to visualize the posterior (or
likelihood) distribution.  It is common then to retrieve
one-dimensional probability by using an integration method
(marginalisation).This technique is mostly used in Bayesian approaches
to parameter estimation. Let's assume that $\vTeta=(x, y, ..., z)$ and
we are interested in plotting the likelihood and finding the 68\%
confidence intervals on $x$, where the other parameters have been
marginalised over. One usually computes:
\begin{eqnarray} 
{\cal L}(x) &=& \int ... \int{\cal L}(d_{obs}|(x,y, ..., z)) P(x,y, ..., z) dy d... dz \label{marg}\\
\int_{0}^{x_m}{\cal L}(d_{obs}|x) dx = 0.5 \;\;\;&&\int_{0}^{x-}{\cal L}(d_{obs}|x) dx = 0.16 \;\;\; \int_{0}^{x+}{\cal L}(d_{obs}|x) dx = 0.84
\end{eqnarray} 
where we assume that $x$ is a positive variable, ${\cal L}$ is
normalized to unity and $P$ is a uniform prior on the
parameters\footnote{The prior is usually taken as uniform in $\vTeta$
in order to show our ignorance on the true value of $\vTeta$, even if
there is no basis in Bayesian theory. In that sense, the interval will
depend on the choice of parameters. Assuming
$\overrightarrow{\gamma}=\vTeta^2$ as parameters and thus a uniform
prior in $\overrightarrow{\gamma}$ will resume in a different interval (see
Fig.~\ref{fig:margmax})}. $[x-, x+]$ is then referred as the 68\%
confidence interval on $x$ with all the other parameters marginalised
over and $x_m$ is quoted as the mean value (such computation of
intervals in referred as EQT for ``equi--probability
tail'')\footnote{Due to the non-linear dependency of the likelihood
against the parameters, the shape of the latter could be highly non
Gaussian. In such cases, it could occur that the maximum of the
likelihood (described earlier as the best model) does not fall inside
the 68\% confidence interval (see for example
Fig.~\ref{fig:margmax}). In that case one should recompute the
interval following the HPD (for ``higher posterior distribution'')
technique, by fixing ${\cal L}(x-)={\cal L}(x+)$ and
$\int_{x-}^{x+}{\cal L}(d_{obs}|x) dx = 0.68$.}. This may be seem
easy in one dimension but could become cumbersome when dealing with 10
dimensional likelihood function (especially for the multi dimensional
integral of the marginalisation Eq.~\ref{marg}).
In order to be less and less dependent of all these effects, and to
decrease the computational time of this step, maximization technique
is mostly used.

\subsection{Maximization}

\begin{figure}[!t]\label{fig:margmax}
\begin{center}
\resizebox{!}{!}{\includegraphics[totalheight=6cm,
        width=6cm]{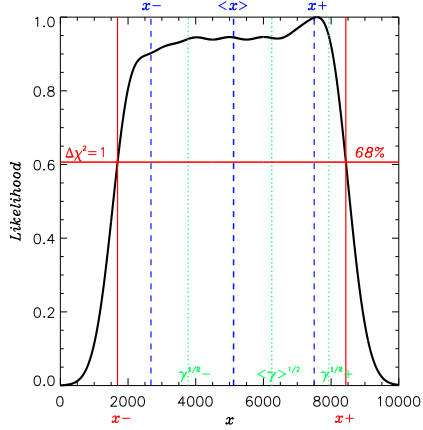}}%\includegraphics[totalheight=6cm,
        %width=6cm]{FIG/allac7_NE.ps}}
%\psfig{figure=maxcomp_ID.ps,width=16cm,height=8cm}
\end{center}
\caption{Comparison between marginalisation and maximization estimation of confidence intervals in an extreme case. The black solid curve is a one-dimensional likelihood function. The blue vertical dashed lines mark the mean and boundaries of 68\% CL interval computed by  integration ($[x-, x+]$). In that case the maximum of the likelihood $\hat x=7580$ is just outside the interval. The green dotted lines are obtained with the same method but with taking $\gamma=x^2$ as variable (see text). Finally the red solid vertical lines shows the interval computed by taking values of the likelihood higher than $exp(-\Delta\chi^2/2)\times{\cal L}_{max}$, where here, ${\cal L}_{max}=1$ and $\Delta\chi^2=1$.}% Right: Contours of likelihood in two dimensions. The shaded regions mark the 2-D confidence intervals. The dashed red lines show the 1-D contours. Once projected, the latter give the confidence intervals on each parameter.}
\end{figure}

For the maximization technique, one assumes also a uniform prior
 in terms of the parameters (typically $P(x,y, ..., z)=1$) but defines
the pdf ($\equiv$ likelihood then) in one dimension as:
\begin{equation} 
{\cal L}(x) = max_{y, ..., z}[{\cal L}(d_{obs}|(x,y, ..., z))] P(x,y, ..., z)
\end{equation} 
which means that for each value of $x$ one takes the maximum of the
likelihood on all the other dimensions. Then, instead of integrating
the resulting one dimensional likelihood like in Eq.~\ref{marg} for
obtaining the confidence intervals, one considers the values of the
likelihood. For example, the boundaries of the 68\% CL region are that
where the likelihood has fallen by a factor $e^{-1/2}$ from its
maximum, ${\cal L}_{max}$. As demonstrated in \cite{tegmark} this
approximation becomes exact  for multivariate Gaussian forms. One
can define different CL intervals by choosing $\Delta \chi^2$ such as
${\cal L}(x_\alpha)/{\cal L}_{max}=e^{-\Delta_\alpha \chi^2 /2}$ where
$\alpha$ marks the confidence level. In one dimension, $\Delta_\alpha
\chi^2 = 1, 4, 9$ for respectively $\alpha = $ 68, 95, 99\%
CL. Fig.~\ref{fig:margmax} shows an example in such a case. 
This technique does not give the real 68, 95 etc. confidence intervals,
obtained only with Monte Carlo simulations by definition, but it is
independent of the choice of the parameter ($x$ versus $x^2$); the
maximum of the likelihood is always inside every interval by
definition, and it is computationally not consuming.  Arguments and
discussion about the different techniques can be found in
\cite{feldman}.

%%%%%%%%%%%%%%%%%%%%%%%%%%%%%%%%%%%%%%%%%%%%%%%%%%%% 5.  practical issues %%

\section{Practical Issues}

We have seen in the previous sections some of the existing statistical
tools needed to perform a proper cosmological parameter estimation.
As one would like to investigate a large number of parameters, and so
a large number of models, some practical issues may be taken in
consideration. I will describe in the following two methods (and some
techniques) which correspond to the two actual ways of determining
cosmological parameters from CMB anisotropies.

\subsection{Cl's computations}

The release of CMBFAST (\cite{cmbfast}) has brought a major improvement
in cosmological parameter estimations. The ability to compute a
theoretical power spectrum in less than one minute (instead of one
hour precedently) has allowed different groups to investigate many
parameters in the same analysis. Different versions of the code have
improved the first release by taking into account many physical
effects (neutrino, reionisation, isocurvature modes, ...)  and
improved the computation by separating small scales effects from
large scales ones (``k-splitting''). Derived from this initial code,
CAMB (\cite{camb}) increased the speed of computation by using FORTRAN
90 facilities. Finally, DASH (\cite{dash}) allows to compute
$C_\ell$'s spectra in few seconds, by interpolating a precomputed grid
of spectra in Fourier space. All these codes are more and more
efficient and fast, and being adapted to be used in parallel
computing.

\subsection{Gridding}

 The gridding method consists in computing the likelihood values of
 different models following a periodical increment for each selected
 parameter, resulting in a $N_{param}$ dimensional
 matrix. Historically, the parameter estimation from CMB anisotropies
 started with small grids of models, typically 3 or 4 free parameters
 with around 10 values each, the other ones fixed to the supposed best
 value of the moment \cite{lineweaver, lasenby}. Then the number of
 parameters increased with the increasing speed of computer processors
 and the development of faster codes to compute the $C_\ell$'s
 (eg. CMBFAST)

 One of the advantages of gridding is that one can compute a grid of
 models, store it and then compute the likelihood with one's set of
 data. If new data come out, one  has just to compute the likelihood
 part again.

 As the number of models investigated increases the storage could
 become a problem \cite{tegmark}. Then, some compression techniques, in
 combination with approximated interpolations, could be applied in
 order to store the necessary information only. The $C_\ell$'s
 computation time may also become a problem. There again approximations
 based on the known behavior of the $C_\ell$'s with parameters have
 been developed \cite{tegmark}.

 One of the inconvenients of the gridding method is that the position
 of the maximum of the likelihood grid is highly dependent of the grid
 itself. Namely, the maximum falls necessarily on one point of the
 grid. This effect is also recurrent when one uses the maximization
 technique. In order to avoid this, spline interpolation techniques
 are used when looking for the maximum along one or more parameters
\cite{tegmark}. 

 Finally, by definition, the gridding method is well adapted to
 multi--processors and data--grid method.

\subsection{Monte Carlo Markov Chains}%%%%%%%%%%%%%%%%%%%%%%% 5.2. Markov %%

 During the last few years, as an alternative to the gridding method,
 the Markov Chain Monte Carlo (MCMC) likelihood analyses had become a
 powerful tool in cosmological parameter estimation. This method
 generates random draws from the posterior distribution that is
 supposed to be a ``realistic'' sample of the likelihood
 hypersurface. The mean, variance, confidence levels can then be
 derived from this sample. Unlike the gridding method, scaling
 exponentially with the number of parameters, the MCMC method scales
 linearly with $N_{param}$ allowing one to explore a larger set of
 parameters or to do the analysis faster.

 Two issues should be highlighted in this method. The first one is the
 step in the random sampling. Typically, the step is taken as the
 standard deviation for each parameter. If it is too large, the chain
 can take a infinite time to converge and the acceptance rate is very
 low. If it is too small the chain will be highly correlated leading
 also to a slow convergence. A second issue is the convergence of the
 chain. At the beginning the sampling of the likelihood is very
 correlated and is not a ``fair'' representation of the posterior
 distribution. After a ``burning period'', the chain converges, the
 samples are independent and the likelihood function could be
 retrieve. The criterium of convergence is not a well defined
 quantity.

 More explanations and applications could be found in
 \cite{mcmc, lewis} and a FORTRAN 90 set of routines is available 
 online \cite{cosmomcmc}.

%%%%%%%%%%%%%%%%%%%%%%%%%%%%%%%%%%%%%%%%%%%%%%%%%%%% 6. Conclusion %%

\section{Conclusions}

In order to derive the cosmological parameters in a given framework
from the temperature fluctuation of the CMB, many steps are
needed. When the observed power spectrum is derived, one could use
different techniques to estimate successively the (approximated)
likelihood value of the family of models (parameters) investigated,
the best model and its goodness of fit, and finally the confidence
intervals on each parameter. Each of these steps may be highly cpu and
memory consuming. With better and better observations, sensitivity and
sky coverage, brute force maximum likelihood methods become
impossible. Many approximations and techniques have then been
developed during the last years, allowing to analyze more and more
data with increasing speed. When the appropriate method is used, this
leads to an unbiased estimate of the cosmological parameters. These
developments have demonstrated that efficient methods could be
developed to take full advantage of data at the Planck accuracy and
allow to determine parameters of cosmological relevance to a
remarkably high accuracy.  This is opening the golden road of
precision cosmology.

%%%%%%%%%%%%%%%%%%%%%%%%%%%%%%%%%%%%%%%%%%%%%%%%%%%%%%%%%%%%
%%%  Acknowledgements  %%%
%%%%%%%%%%%%%%%%%%%%%%%%%%
\Acknowledgements{MD would like to thank A. Blanchard, J. Bartlett and K. Moodley for useful discussions and corrections. MD acknowledge an EU CMB-Network fellowship}

%%%%%%%%%%%%%%%%%%%%%%%%%%%%%%%%%%%%%%%%%%%%%%%%%%%%%%%%%%%%
%%%  Bibliography  %%%
%%%%%%%%%%%%%%%%%%%%%%%

%
\end{document}